\documentclass[%
15pt,
reprint,
 superscriptaddress,
 amsmath,amssymb,
 aps,UTF8,
]{revtex4-1}

\usepackage{graphicx}
\usepackage{dcolumn}
\usepackage{bm}
\usepackage{xcolor}
\usepackage{url}
\usepackage{float}
\usepackage{tabularx}
\usepackage{lipsum}
\usepackage{array}

\usepackage{booktabs}

\usepackage{subfigure}
\usepackage{multirow}

\begin{document}


\title{Exploring Short-Range Correlations in Symmetric Nuclei: Insights into Contacts and Entanglement Entropy}
\author{Wei Kou}
\email{kouwei@impcas.ac.cn}
\affiliation{Institute of Modern Physics, Chinese Academy of Sciences, Lanzhou 730000, China}
\affiliation{School of Nuclear Science and Technology, University of Chinese Academy of Sciences, Beijing 100049, China}

\author{Jingxuan Chen}
\affiliation{Institute of Modern Physics, Chinese Academy of Sciences, Lanzhou 730000, China}
\affiliation{Guangdong Provincial Key Laboratory of Nuclear Science, Institute of Quantum Matter, South China Normal University, Guangzhou 510006, China}

%

%
%
\author{Xurong Chen}
\email{xchen@impcas.ac.cn (Correspongding author)}
\affiliation{Institute of Modern Physics, Chinese Academy of Sciences, Lanzhou 730000, China}
\affiliation{School of Nuclear Science and Technology, University of Chinese Academy of Sciences, Beijing 100049, China}



\begin{abstract}

The Short-Range Correlations between nucleons in nuclei is regarded as a complex system. We investigate the relationship between the orbital entanglement entropy of SRCs $S_{ij}$ in nuclear structures and Tan contact $c_{ij}$, and find that the orbital entanglement entropies and Tan contacts corresponding to proton-proton SRC pairs and neutron-proton SRC pairs in nuclei demonstrate a scaling relation. More specifically, the proportionality of entanglement entropy between proton-proton pairs and neutron-proton pairs is directly related to the ratio of nuclear contacts within the atomic nucleus, demonstrating an approximate ratio of 2.0. Our research suggests that this scaling relationship should hold true for all symmetric nuclei, furthermore, we offer a possible explanation for this phenomenon.
\end{abstract}

\maketitle


\section{Introduction}
\label{sec:intro}
The atomic nuclei are complex and strongly interacting systems that are difficult to solve exactly. As an approximation, the nuclear forces can be separated into long-range attractive as well as short-range repulsive interactions. Strong attractive forces and strong repulsive forces between two, three or even more nuclei can produce a complicated system and form a unique ground state of the nucleus, a phenomenon known as Short-Range Correlation (SRC) \cite{Frankfurt:1988nt,Frankfurt:2008zv,CiofidegliAtti:2015lcu}. The details of the SRC effect are important inspiration for understanding and studying topics such as symmetry energies of nuclear matter \cite{Frankfurt:2008zv,Hen:2014yfa,Cai:2015xga,Hen:2016ysx}, mergers of neutron stars \cite{Baym:2017whm}, and lepton-nucleus scattering processes \cite{Accardi:2012qut,Anderle:2021wcy}. For some review articles on SRC physics please refer to \cite{Arrington:2011xs,Hen:2013oha,Hen:2016kwk,Fomin:2017ydn,Arrington:2022sov}. Experimentally investigating SRC employs electron (nucleon)-nucleus scattering process with high-energy and large-momentum transfer \cite{Frankfurt:1993sp,Tang:2002ww,CLAS:2003eih,CLAS:2005ola,Piasetzky:2006ai,JeffersonLabHallA:2007lly,Subedi:2008zz,CLAS:2010yvl,Fomin:2011ng,Hen:2014nza,LabHallA:2014wqo,CLAS:2018xvc,JeffersonLabHallA:2020wrr,Li:2022fhh}. A large number of measurements have shown that SRC pairs within the nucleus are dominated by the deuteron form, i.e., neutron-proton ($np$) pairs. Quantitatively described as $np$ pairs are about 20 times more numerous than the other channels ($pp$ or $nn$) \cite{Subedi:2008zz,LabHallA:2014wqo,CLAS:2018xvc}. Meanwhile, with ab-initio calculations \cite{CiofidegliAtti:1991mm,CiofidegliAtti:1995qe,Schiavilla:2006xx,Alvioli:2007zz,Feldmeier:2011qy,Alvioli:2012qa,Rios:2013zqa,CiofidegliAtti:2017xtx,CiofidegliAtti:2017tnm}, there are more progresses have been discussed on SRCs. 

The mean-field approximation enables the examination of the global nature of atomic nuclei. However, due to the significant momentum transfer associated with SRCs, the study of dilute nuclear matter becomes unfeasible. Extensive research on atomic nuclei, employing simple Fermi momentum distributions and cold atomic gas methods, has revealed the inadequate ability to accurately describe the behavior of SRCs. Consequently, a more sophisticated approach is required to elucidate the interactions between nucleons in SRC states. Recently, the investigation of nuclear physics has incorporated the concepts of entropy in thermodynamics and information science, leading to a burgeoning research area. For a comprehensive list of recent references, please refer to \cite{Ehlers:2022oal}. Notably, the use of information entropy to investigate nuclear structure has yielded several innovative ideas. The application of quantum entanglement entropy, a specific form of information entropy, is widely utilized across various disciplines \cite{Kharzeev:2017qzs,Kharzeev:2021nzh,Kharzeev:2021yyf,Wang:2016sfq,Kou:2022dkw,Kou:2023azd,Hentschinski:2021aux,Hentschinski:2022rsa}. The interplay between entanglement entropy and SRCs has been explored in multiple studies. Specifically, Refs. \cite{Bulgac:2021hbx,Bulgac:2022cjg,Bulgac:2022yyu,Pazy:2022mmg,Bulgac:2022ygo} offer insights into the relationship between entanglement entropy and SRCs.

As mentioned in Ref. \cite{Kou:2022dkw}, entanglement is a fundamental feature of quantum systems. In principle, it applies to any quantum pure state which can be divided into different subsystems.  Being SRCs states of nuclear many-body systems, they should generate entanglement at any energy. Calculating the entanglement entropy of SRCs, i.e., quantizing the degree of entanglement is required to study the quantum entanglement states of SRCs. The entanglement entropy of the nuclear structure is the first to be considered. In recent studies the entanglement entropy of nuclei such as $^4$He, $^6$He, etc. have been discussed in \cite{Robin:2020aeh}. In addition, the discussion of single-orbital or two-orbital mutual information for nuclei such as $^{28}$Si, $^{56}$Ni and $^{64}$Ge using density matrix renormalization group (DMRG) schemes have also been addressed in recent work \cite{Legeza:2015fja}. The investigation of the entanglement entropy of the SRC is essentially a scaling separation of the eigenstates of the nucleus, i.e., the SRC nuclear states for high momentum orbital and the mean-field approximation of the Fermi momentum distribution nucleon states for low-momentum orbital. One approach to addressing scaling separation is called the Generalized Contact Formalism (GCF) \cite{Weiss:2014gua,Weiss:2015mba,Weiss:2016obx,Weiss:2017huz}. The nuclear contacts were defined which inspired by Tan contact theory in atomic physics \cite{Tan_2008f,Tan_2008s,Tan_2008t}. The application of the GCF method to describe the fundamental properties of the SRCs have achieved some successes, such as two-body nucleon densities \cite{Weiss:2017huz}, high-momentum tails \cite{Weiss:2015mba,Weiss:2017huz,Alvioli:2016wwp}, and electron-nucleus scattering experiments \cite{Weiss:2018tbu}. The connection between nuclear contacts and SRCs information science was also discussed recently \cite{Robin:2020aeh,Bulgac:2022ygo,Pazy:2022mmg}.

In this work, we find the scaling relation of nuclear contact and entanglement entropy in SRCs nucleus. Through our viewpoint, the extraction and analysis of existing nuclear contacts and the corresponding entanglement entropy satisfy a scaling relation among them. One should conclude that the relationship between nuclear contacts and the single-orbital entanglement entropy constructed from them - a relationship that should apply to nuclei of any mass number $A$ and may predict the existence of SRC channel ratios ($pp/np$) for nuclei that have not been measured yet. We start with simple review of GCF theory and single-orbital entanglement entropy in Sec. \ref{sec:formalism}. In Sec. \ref{sec:resluts and discussion}, we present and show our calculations and main results about nuclear contacts and scaling law of SRCs. Meanwhile, some discussion are given. Finally, we conclude our work and give the outlook. We emphasize here that for the convenience of the whole discussion only the symmetric nuclei case is considered, and that the relevant corrections for asymmetric nuclei case can be found in Ref. \cite{Weiss:2018zrd}.

\section{Formalism}
\label{sec:formalism}

\subsection{General contact formalism}
\label{subsec:GCF}
The GCF method is a decomposition of nuclear many-body wave function into spatially close correlated nucleon pair two-body part and other components. If the correlated nucleon pair is chosen to be the universal state, the remainders of them imply situation-dependent state consisting of $A-2$ nucleons. Therefore, the factorized asymptotic wave-function takes the form \cite{Weiss:2015mba}
\begin{equation}
	\Psi\xrightarrow{r_{ij}\to0}\sum_\alpha\varphi_\alpha(\boldsymbol{r}_{ij})A_{ij}^\alpha(\boldsymbol{R}_{ij},\{\boldsymbol{r}\}_{k\neq ij}),
	\label{eq:GCF1}
	\end{equation}
	where the index $ij$ corresponds to $np$, $pp$, and $nn$ pairs. The pair wave functions depend on the total spin of the pair $s$, and its angular momentum quantum number $l$ (with respect to the relative coordinate $\boldsymbol{r}_{ij}$) which are coupled to create the total pair angular momentum $j$. The quantum numbers $\alpha=s,\ l,\ j$ define the pair's channel. The parts of $\varphi_\alpha(\boldsymbol{r}_{ij})$ are the two-body universal functions defining the SRC state, $A_{ij}^\alpha$ denote the so called regular parts of the  many-body wave function, the index $\alpha$ represents the quantum numbers of two-body states. The $\varphi_\alpha(\boldsymbol{r}_{ij})$ is a function of the distance $\boldsymbol{r}_{ij}$ between nucleon pair with SRC states rather than of the center-of-mass system coordinate $\boldsymbol{R}_{ij}$ appearing in $A_{ij}^\alpha$. The later one is obtained by solving the two-body, zero energy, Schrödinger equation with the full many-body potential.
	
	Under the approximation discussed above, the nuclear contact of GCF is simply defined as \cite{Weiss:2015mba}
	\begin{equation}
		C_{ij}^\alpha=16\pi^2N_{ij}(A,Z)\langle A_{ij}^\alpha|A_{ij}^\alpha\rangle,
		\label{eq:nuclear contact}
	\end{equation}
	Since we are interested in the symmetric nuclei case, the $N$ as a nucleon pairs number is determined by $Z$ protons and $A-Z$ neutrons, one can consider $Z=A/2$.

	If we return to the SRC orbitals, i.e., $k_F\ll|\boldsymbol{k}|$, the momentum distribution can be approximated by GCF theory \cite{Weiss:2016obx}
	\begin{equation}
	\begin{aligned}
		n_p(\boldsymbol{k})&= 2{C}_{pp}^{s=0}|{\phi}_{pp}^{s=0}(\boldsymbol{k})|^2+{C}_{pn}^{s=0}|\phi_{pn}^{s=0}(\boldsymbol{k})|^2  \\
		&+C_{pn}^{s=1}|\phi_{pn}^{s=1}(\boldsymbol{k})|^2
	\end{aligned}
		\label{eq:one distri2}
	\end{equation}
	where $\phi^\alpha(\boldsymbol{k})$ is the Fourier transform of function $\varphi_\alpha(\boldsymbol{r}_{ij})$ and the same when replacing $n$ with $p$ in (\ref{eq:one distri2}) \cite{Weiss:2015mba}. According to normalisation condition $\int_{k_F}^{\infty} |\phi(\boldsymbol{k})|^2\mathrm{d} \boldsymbol{k}=1$, the fraction of the one-body momentum density with the momentum above $k_F$ is given by \cite{Weiss:2016obx}
	\begin{equation}
		\frac{\int_{k_F}^\infty n(\boldsymbol{k})d\boldsymbol{k}}{\int_0^\infty n(\boldsymbol{k})d\boldsymbol{k}}=\frac{C_{nn}^{s=0}+C_{pp}^{s=0}+C_{np}^{s=0}+C_{np}^{s=1}}{A/2}.
		\label{eq:GCF-frac}
	\end{equation}
	where  $n(\boldsymbol{k})=n_n(\boldsymbol{k})+n_p(\boldsymbol{k})$. Note that we consider the contribution of the main channels of the SRCs, e.g., the $np$ deuteron channel $(l=0,2$ and $s=1$ coupled to $j=1$), and the singlet $pp$, $np$, and $nn$ $s$-wave channel ($l=s=j=0$). $C_{ij}^s/\frac{A}{2}$ gives the fraction of the one-body momentum density above the Fermi momentum due to each type of SRC pair \cite{Weiss:2016obx}. In fact, the above GCF has been successful in explaining the one-body as well as two-body density distributions of nucleons \cite{Weiss:2016obx,Weiss:2018zrd}.

\subsection{Single-orbital entanglement entropy}
\label{subsec:EE}
The origin of entanglement entropy is distinct from the conventional notion of entropy attributed to a lack of knowledge concerning the microstate of a system, originating from thermal fluctuations. Rather, entanglement entropy stems from the intricate entanglement prevailing among distinct subunits of the system \cite{Plenio:2007zz,Horodecki:2009zz}. In order to consider the scaling separation of SRC nuclei \cite{Tropiano:2021qgf}, a simple model is to introduce orbital entanglement entropy. The simplified model in which the SRC is identified with the high-momentum subspace and considered as a single orbital. Thus a nucleon can occupy either one of the Fermi sea (FS) orbitals or SRC. In this way, the Hilbert space of nucleus can be divided into the tensor product of the FS and the SRC space orbitals
\begin{equation}
	\mathcal{H}=\mathcal{H}_\mathrm{FS} \otimes\mathcal{H}_\mathrm{SRC}.
	\label{eq:Hilbert}
\end{equation}
It is a product of a FS Hilbert subspace $\mathcal{H}_\mathrm{FS}$ and the SRC Hilbert subspace $\mathcal{H}_\mathrm{SRC}$.

We use establishment process of the single-orbital entanglement entropy in Ref. \cite{Robin:2020aeh}, which essentially yields the reduced density matrix of subsystems. The nucleus eigenstates can then be written as a linear combination of Slater determinants $|\phi\rangle$ for the nucleon wave functions,
\begin{equation}
	|\Psi\rangle=\sum_\eta\mathcal{A}_\eta|\phi_\eta\rangle,
	\label{eq:slater1}
\end{equation}
where the Slater determinant is given in terms of applying creation operators on the real particle vacuum $|0\rangle$:
\begin{equation}
	|\phi_\eta\rangle=\prod_{i\in\eta}^Aa_i^\dagger|0\rangle,
	\label{eq:Slater2}
\end{equation}
where $A$ is the nucleus mass number.

According to this way, the single-orbital reduced density matrix is \cite{Robin:2020aeh}
\begin{equation}
	\rho_{n_i,n_i^{\prime}}^{(i)}=\sum_{BC}\left<\Psi|BC\right>\left|n_i^{\prime}\right>\left<n_i\right|\left<BC|\Psi\right>,
	\label{eq:reduced dm}
\end{equation}
where $BC=n_1n_2,\cdots n_i,n_{i+1},\cdots n_A$. Each state $i$ has the possibility of being occupied or empty. The basis $\{|n_i\rangle\}$ denotes $\{|0\rangle,|1\rangle=a_i^\dagger|0\rangle\}$. With this basis the density matrix is written as \cite{Robin:2020aeh,Pazy:2022mmg}
\begin{equation}
\rho^{(i)}=\left(\begin{matrix}1-\gamma_{ii}&0\\0&\gamma_{ii}\end{matrix}\right),
\label{eq:matrix}
\end{equation}
where the occupation of the orbital is given by $\gamma_{ii}=\langle \Psi|a_i^\dagger a_i |\Psi\rangle$.  Thus, one can construct the von Neumann entropy from the density matrix (\ref{eq:matrix})
\begin{equation}
	S_i^{(1)}=-\mathrm{Tr}[\rho^{(i)}\ln\rho^{(i)}]=-\sum_{k=1}^2\omega_k^{(i)}\ln\omega_k^{(i)},
	\label{eq:EE}
\end{equation}
where $\omega_k^{(i)}$ is the eigenvalue of $\rho^{(i)}$. Here we emphasize that Eq. (\ref{eq:EE}) is an expression for the single orbital entanglement entropy, and the corresponding density matrix is of $2\times 2$ form.

\subsection{Single-orbital entanglement entropy with nuclear contact}

At present, our discussion can be succinctly summarized as follows: Firstly, the atomic nucleus system can be divided into two distinct scales, namely, SRC orbitals with momentum exceeding the Fermi momentum, and FS orbitals with momentum below the Fermi momentum. These two types of orbitals are quantum entangled. Secondly, nuclear contacts can be constructed using the GCF method. Thirdly, the density matrix of entanglement entropy for the SRC  single orbital is correlated to the occupancy of the orbital. In the following, we provide a brief description of how the SRC single-orbital entanglement entropy can be represented in terms of nuclear contacts.

Since SRC is characterized by a high momentum tail compared with FS, one can consider the nucleons with the momenta $k>k_F$ as occupying high momentum SRC orbitals. According to the GCF, Eq. (\ref{eq:GCF-frac}) represents the ratio of high-momentum orbital nucleons to total nucleons. If one defines the operator $\hat{P}=a_k^\dagger a_k|_{k>k_F}$, the probability that a nucleon in a given nucleus occupies an SRC orbital can be easily obtained as
\begin{equation}
	\gamma_{\mathrm{SRC}}=\langle\Psi|\hat{P}|\Psi\rangle=\frac{C_{ij}^\alpha}{A/2}= c_{ij}^\alpha,
	\label{eq:reduced C}
\end{equation}
where $c_{ij}^\alpha$ is the normalised (reduced) nuclear contact and can be extracted by nucleons two-body wave function and momentum distribution \cite{Weiss:2016obx}. According to the above definition of SRC occupancy probability, the single orbital entanglement entropy entropy for a single SRC is directly obtained through Eq. (\ref{eq:matrix})
\begin{equation}
       S^{\mathrm{SRC}}(c_{ij}^\alpha)=-\biggl[c_{ij}^\alpha\ln\left(\frac {c_{ij}^\alpha}{1-c_{ij}^\alpha}\right)+\ln\left(1-c_{ij}^\alpha\right)\biggr].
       \label{eq:SEE}
\end{equation}
To obtain the total SRC orbital entanglement entropy one has to multiply the single SRC entanglement by nucleon pair number $N(A,Z)=A/2$
\begin{equation}
	S^{\mathrm{SRC}}_{tot}(A,c)=-\frac{A}{2}\biggl[c_{ij}^\alpha\ln\left(\frac{c_{ij}^\alpha}{1-c_{ij}^\alpha}\right)+\ln\left(1-c_{ij}^\alpha\right)\biggr].
	\label{eq:SEE2}
\end{equation}
This expression reveals the linear dependence of the entanglement entropy on the nuclear mass number $A$. In other words the total SRC entanglement entropy is proportional to the volume of the nucleus. This easily piques people's curiosity, as the entanglement entropy is frequently associated with the system's area law -- the Bekenstein-Hawking entropy \cite{Bekenstein:1973ur,Hawking:1974rv,Hawking:1975vcx}.

\section{Results and discussions}
\label{sec:resluts and discussion}
We review the formalism for computing the entanglement entropy of SRC orbitals in Section \ref{sec:formalism}. An example of calculating entanglement entropy is given in Ref. \cite{Pazy:2022mmg}, however the absolute magnitude of entanglement entropy is not our focus in this work. Since there is experimental interest in the ratio of SRC nucleon pair types in nuclei, i.e., the ratio of the number of proton-proton pairs in the SRC state to the number of neutron-proton pairs in a given nucleus. From the perspective of nuclear contacts it seems possible to qualitatively use the ratio of the reduced contacts of the corresponding channel as a basis for determining this ratio \cite{Weiss:2016obx}. In this section we start from the relation between the reduced nuclear contact and the entanglement entropy to discuss what relations should be satisfied by the ratios between the different SRC channels in the nucleus.

In fact, it is viable to extract nuclear contacts, although the nuclear many-body wavefunction cannot be fully solved, some practical methods are given in Ref. \cite{Weiss:2016obx}. The authors argued that the nucleons two-body functions used in their work were calculated numerically using the AV18 potential for zero energy \cite{Weiss:2018zrd}. The obtained wave functions are insensitive to the exact value of the energy for small distances and large momenta. In Ref. \cite{Weiss:2016obx}, the authors have used three methods for the extraction of nuclear contacts, employing two-body density distributions \cite{Wiringa:2013ala} in coordinate space and momentum space as the first two methods, respectively. In the third method, they used experimental data \cite{Subedi:2008zz,LabHallA:2014wqo,Hen:2014nza,JeffersonLabHallA:2007lly}.

The extracted nuclear contacts are shown in Table 1 of Ref. \cite{Weiss:2016obx}. We consider all the extraction results for the symmetric nuclei case to compute the corresponding SRC orbital entanglement entropies. In this work, we are only responsible for the entropy between the different SRC channels and the ratio of the nuclear contacts. Using Eq. (\ref{eq:SEE}) one should get the expression which describes the ratio of SRC entanglement entropies and reduced nuclear contacts with different channels
\begin{equation}
	R(c_{pp},c_{np})=\frac{S_{pp}^{SRC}/S_{np}^{SRC}}{c_{pp}/c_{np}},
	\label{eq:ratio}
	\end{equation}
	where the $pp$-channel we consider the contribution of spin 0, and the $np$-channel we consider the total contribution of spin 0 and 1. The nuclear contacts of the symmetric nuclei were extracted in Ref. \cite{Weiss:2016obx} and the ratios defined by Eq. (\ref{eq:ratio}) are shown in Table. \ref{tab:contacts}.

\begin{table*}[htbp]
	\centering
	\caption{The nuclear contacts and the corresponding ratios defined in Eq. (\ref{eq:ratio}) for a variety of nuclei. The contacts come from Ref. 
	\cite{Weiss:2016obx}, which are divided by $A/2$ and give the percent of nucleons above Fermi energy $k_F$ in the different SRC channels. Only the symmetric nuclei case is taken into account, and we find that the ratios computed from the reduced contacts given by either $k$-space or $r$-space converge almost to 2.}
	\label{tab:contacts}
		\begin{tabular}{|c|ccc|ccc|}
			\hline
			\multirow{2}{*}{\textbf{$A$}} & \multicolumn{3}{c|}{\textbf{k-space}}                                                                                            & \multicolumn{3}{c|}{\textbf{r-space}}                                                                                                                                \\ \cline{2-7} 
			& \multicolumn{1}{c|}{\textbf{$c_{np}$}} & \multicolumn{1}{c|}{\textbf{$c_{pp}$}} & \textbf{$R=\frac{S_{pp}/S_{np}}{c_{pp}/c_{np}}$} & \multicolumn{1}{c|}{\textbf{$c_{np}$}}                  & \multicolumn{1}{c|}{\textbf{$c_{pp}$}}                    & \textbf{$R=\frac{S_{pp}/S_{np}}{c_{pp}/c_{np}}$} \\ \hline
			\textbf{$^4$He}               & \multicolumn{1}{c|}{0.1299$\pm$0.0010} & \multicolumn{1}{c|}{0.0065$\pm$0.0003} & 2.029$\pm$0.167                                & \multicolumn{1}{c|}{\multirow{2}{*}{0.1218$\pm$0.0003}} & \multicolumn{1}{c|}{\multirow{2}{*}{0.00567$\pm$0.00004}} & \multirow{2}{*}{2.028$\pm$0.003}               \\ \cline{1-4}
			\textbf{$^4$He (exp)}         & \multicolumn{1}{c|}{0.157$\pm$0.007}   & \multicolumn{1}{c|}{0.008$\pm$0.002}   & 2.104$\pm$0.098                                & \multicolumn{1}{c|}{}                                   & \multicolumn{1}{c|}{}                                     &                                                \\ \hline
			\textbf{$^6$Li}               & \multicolumn{1}{c|}{0.1103$\pm$0.0011} & \multicolumn{1}{c|}{0.0049$\pm$0.0003} & 2.007$\pm$0.021                                & \multicolumn{1}{c|}{0.1056$\pm$0.0004}                  & \multicolumn{1}{c|}{0.00415$\pm$0.00004}                  & 2.030$\pm$0.004                                \\ \hline
			\textbf{$^{8}$Be}             & \multicolumn{1}{c|}{0.1406$\pm$0.0022} & \multicolumn{1}{c|}{0.0079$\pm$0.0007} & 2.021$\pm$0.033                                & \multicolumn{1}{c|}{0.126$\pm$0.001}                    & \multicolumn{1}{c|}{0.00603$\pm$0.00003}                  & 2.032$\pm$0.006                                \\ \hline
			\textbf{$^{10}$B}             & \multicolumn{1}{c|}{0.1259$\pm$0.0022} & \multicolumn{1}{c|}{0.0079$\pm$0.0006} & 1.941$\pm$0.028                                & \multicolumn{1}{c|}{0.1127$\pm$0.0020}                  & \multicolumn{1}{c|}{0.0057$\pm$0.0002}                    & 1.973$\pm$0.016                                \\ \hline
			\textbf{$^{12}$C}             & \multicolumn{1}{c|}{0.182$\pm$0.008}   & \multicolumn{1}{c|}{0.013$\pm$0.002}   & 2.047$\pm$0.071                                & \multicolumn{1}{c|}{\multirow{2}{*}{0.1573$\pm$0.0010}} & \multicolumn{1}{c|}{\multirow{2}{*}{0.0083$\pm$0.0001}}   & \multirow{2}{*}{2.092$\pm$0.007}               \\ \cline{1-4}
			\textbf{$^{12}$C (exp)}       & \multicolumn{1}{c|}{0.195$\pm$0.021}   & \multicolumn{1}{c|}{0.015$\pm$0.005}   & 2.052$\pm0.163$                                & \multicolumn{1}{c|}{}                                   & \multicolumn{1}{c|}{}                                     &                                                \\ \hline
			\textbf{$^{16}$O}             & \multicolumn{3}{c|}{}                                                                                                            & \multicolumn{1}{c|}{0.1208$\pm$0.0030}                  & \multicolumn{1}{c|}{0.0068$\pm$0.0003}                    & 1.963$\pm$0.022                                \\ \hline
			\textbf{$^{40}$Ca}            & \multicolumn{3}{c|}{}                                                                                                            & \multicolumn{1}{c|}{0.1233$\pm$0.0030}                  & \multicolumn{1}{c|}{0.0073$\pm$0.0004}                    & 1.953$\pm$0.025                                \\ \hline
		\end{tabular}%
\end{table*}

	\begin{figure*}[htbp]
	\centering
	\subfigure{
		\includegraphics[width=0.46\textwidth]{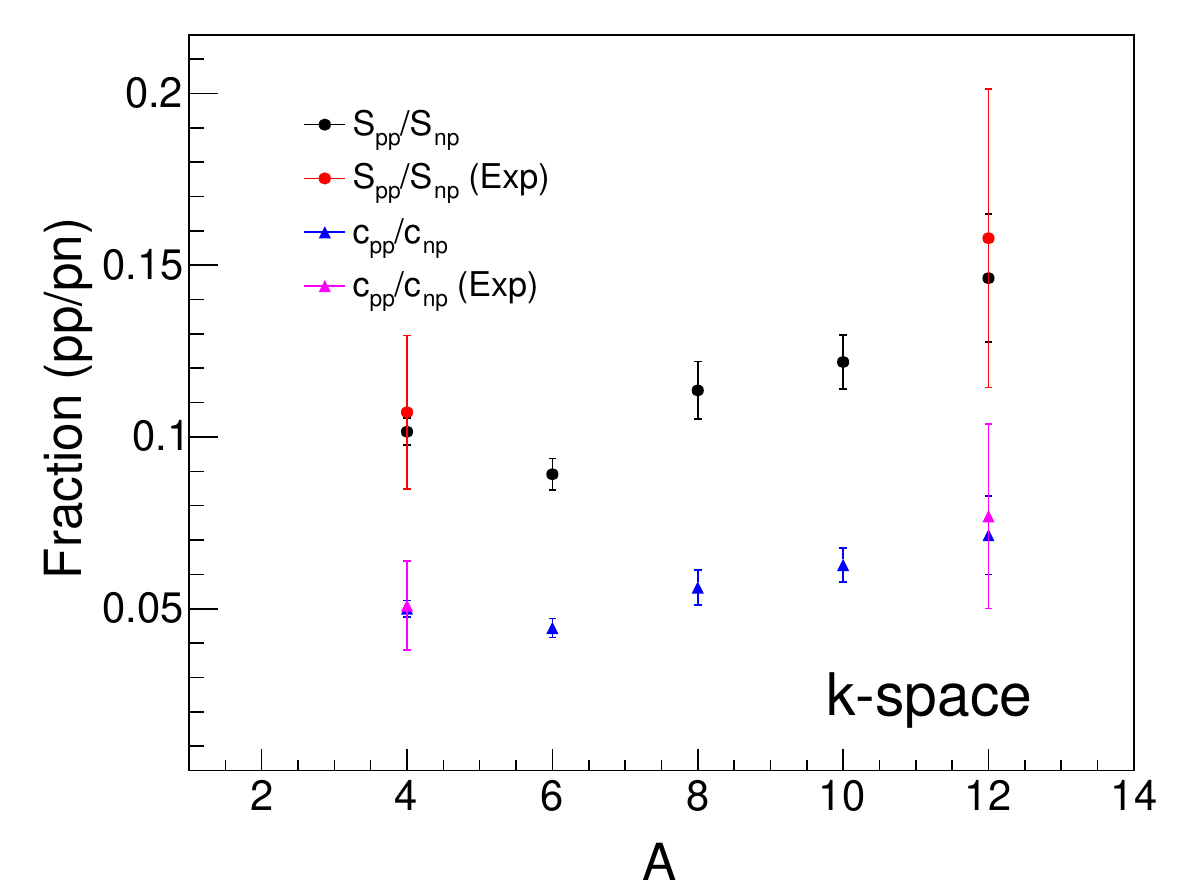}}
	\subfigure{
		\includegraphics[width=0.46\textwidth]{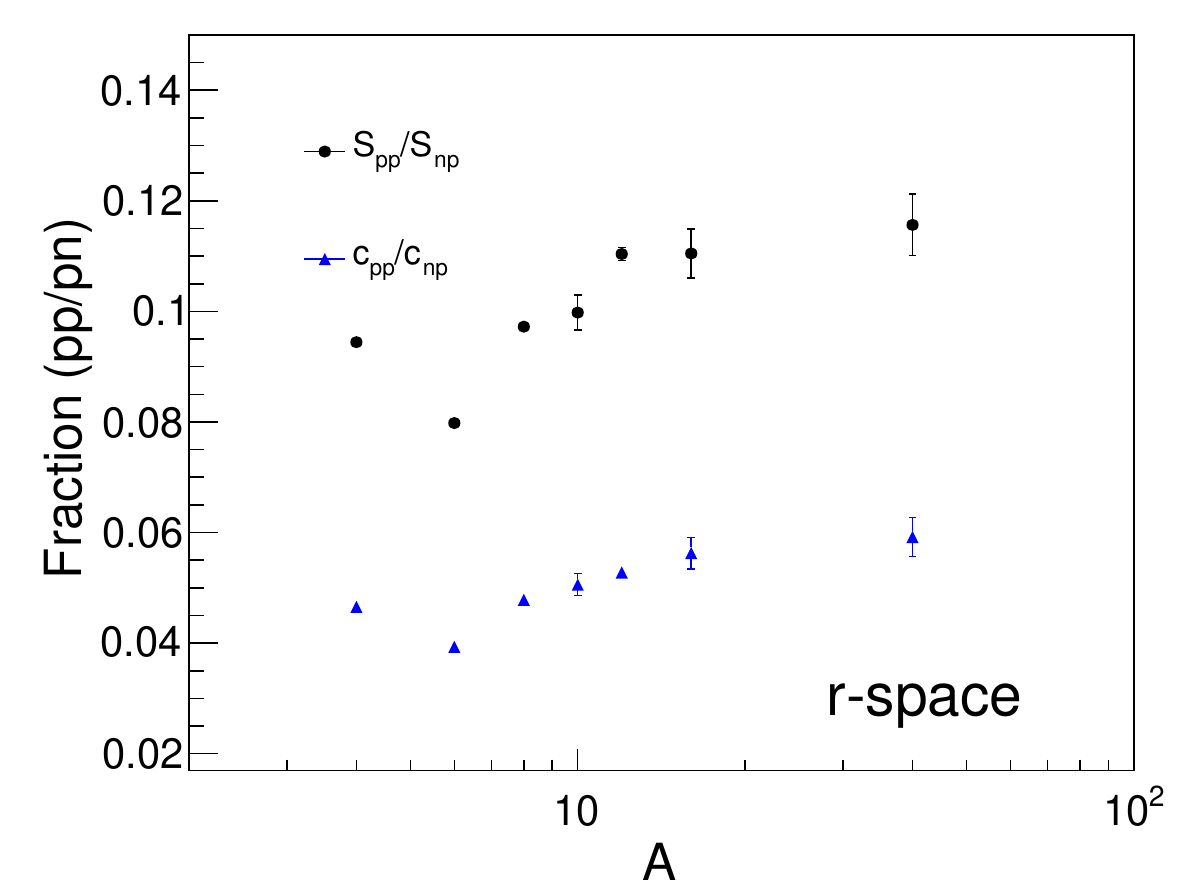}}
	\caption{The ratio between the nuclear contacts of proton-proton and neutron-proton channels and the ratio corresponding SRC single-orbital entanglement entropies. Nuclear contacts are extracted from the form of the two-body wave function in $k$-space (left) and $r$-space (right) mentioned in Ref. \cite{Weiss:2016obx}.}
	\label{fig: ratio-B}
\end{figure*}

\begin{figure*}[htbp]
	\centering
	\subfigure{
		\includegraphics[width=0.46\textwidth]{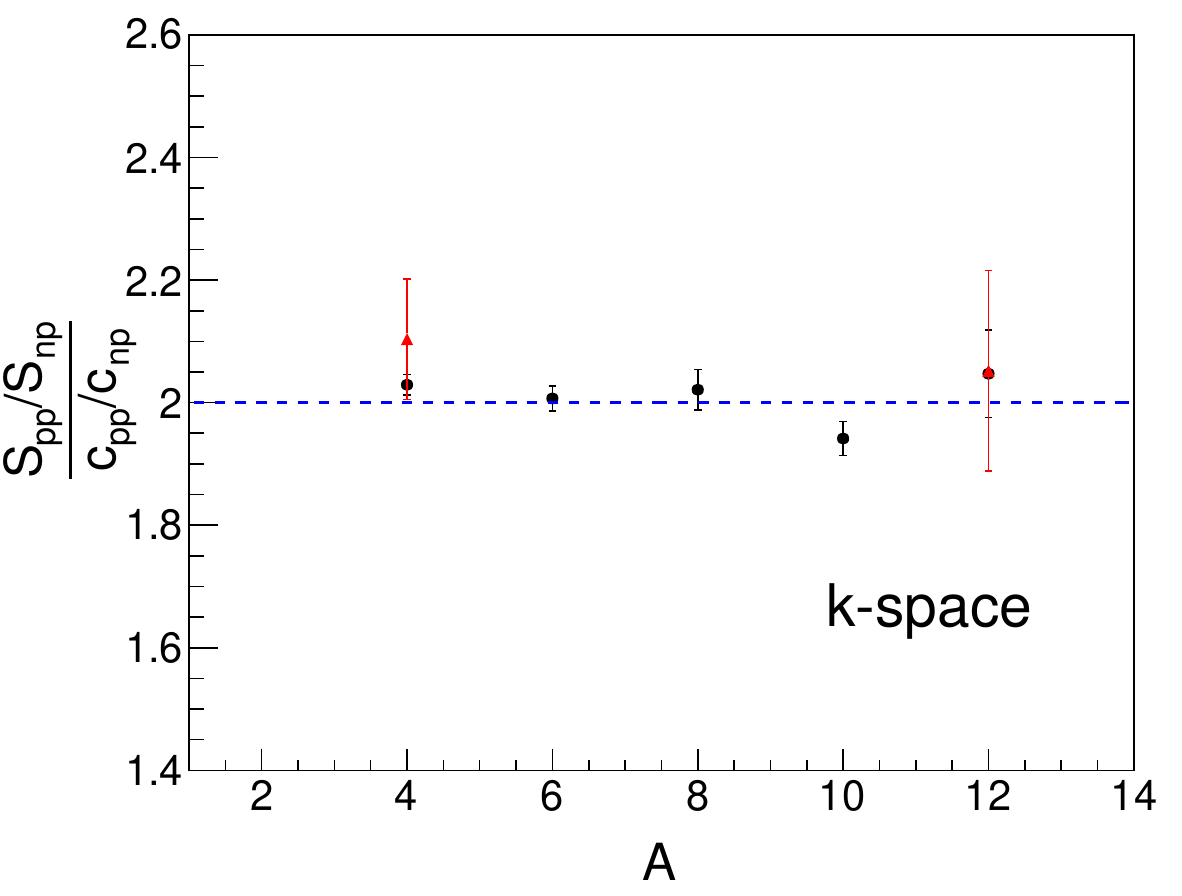}}
	\subfigure{
		\includegraphics[width=0.46\textwidth]{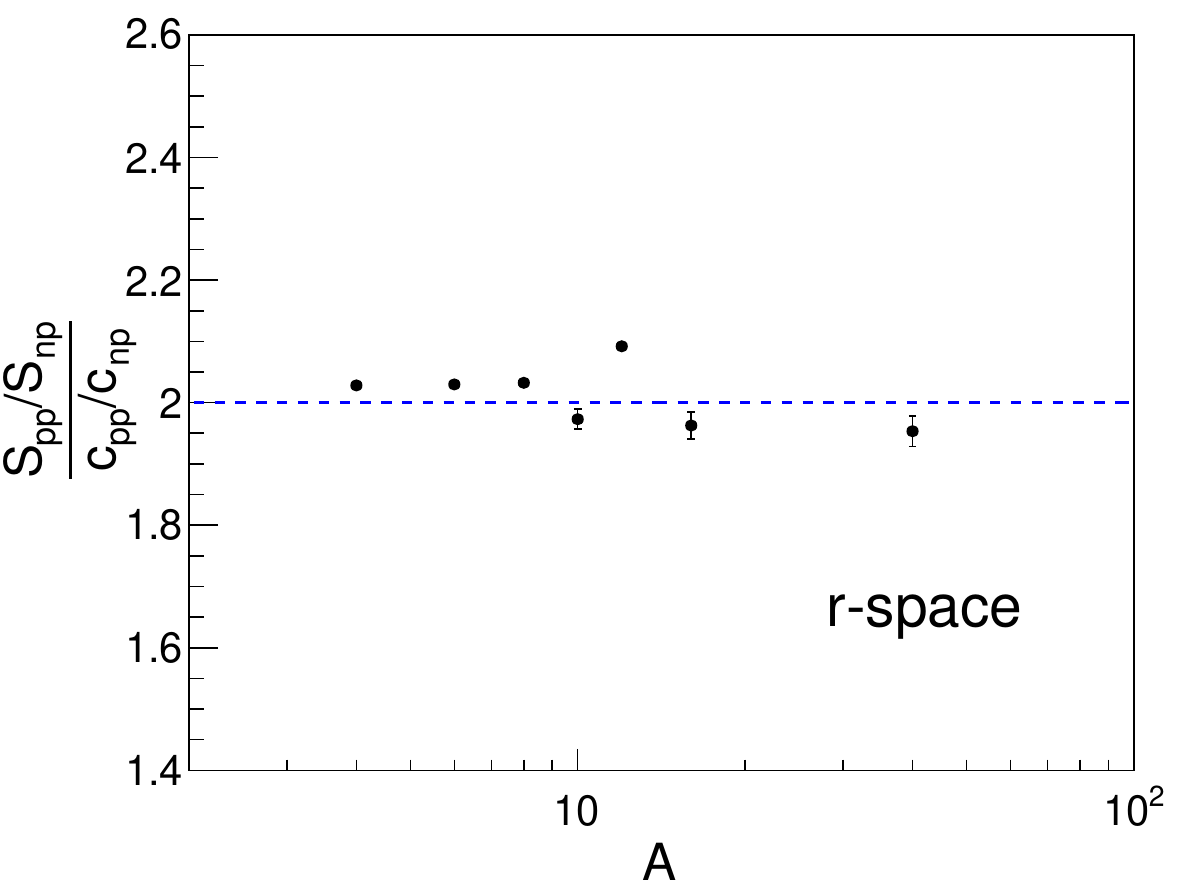}}
		\caption{Left: Calculated ratios from reduced nuclear contacts extracted from the $k$-space nuclear two-body density distribution versus the nuclei mass numbers, where the red data points indicate the results obtained from the experiments \cite{Subedi:2008zz,LabHallA:2014wqo,Hen:2014nza,JeffersonLabHallA:2007lly,Weiss:2016obx}. Right: Calculated ratios from reduced nuclear contacts extracted from the $r$-space nuclear two-body density distribution versus the nuclei mass numbers. The blue dashed line represents a value of 2 on the vertical axis.}
		\label{fig: ratio-A}
\end{figure*}

We show the specific ratio relationships in Figures .\ref{fig: ratio-B} and \ref{fig: ratio-A}, corresponding to the extracted contacts from $k$-space and $r$-space , respectively. The vertical axis in the Figure. \ref{fig: ratio-A} represents the corresponding ratio of Eq. (\ref{eq:ratio}). The source of uncertainties is taken from the uncertainties of the nuclear contacts extracted from Ref. \cite{Weiss:2016obx}. From Table. \ref{tab:contacts} and Figures. \ref{fig: ratio-B} and \ref{fig: ratio-A} one should find that the value of Eq. (\ref{eq:ratio}) barely depends on the nucleon number $A$ of the atomic nucleus and converges to a constant -- 2. However, considering the ratio of entanglement entropy or the ratio of nuclear contacts alone does not seem to determine that they are mass number dependent. This is a very specific result, discussing separately the ratios of nuclear contacts for different channels and the raitos of SRC entanglement entropy for different channels have no unambiguous convergence behavior. Next we try to analyze and discuss this phenomena.

Generally speaking, there are at least two possibilities for making $R(c_{pp},c_{np})=constant$ in Eq. (\ref{eq:ratio}) to hold: the numerator and denominator are each constants; neither the numerator nor the denominator is constant, but the total ratio implies some kind of constant relation. Both possibilities are related to nuclear contacts. The question of how to obtain the nuclear contacts of a symmetric nucleus is the focus of our discussion, which conceptually requires a two-body density distribution of nucleon pairs. To simplify the discussion, we consider the description of nuclear contacts in Ref. \cite{Weiss:2018zrd}. The authors used the two-body nucleon charge density to construct nuclear contacts instead of the nuclear two-body density and came to the following conclusions \cite{Weiss:2018zrd}:
\begin{equation}
	\begin{aligned}
	&C_{pp}^{s=0}=C_{np}^{s=0}\approx\frac9{40\pi}\frac1{R_0^3}\frac1{|\varphi_{pp}^{s=0}(r_0)|^2}\frac{Z^2}A,\\
	&C_{pn}^{s=1}=L\frac{NZ}AC_{pn}^{s=1}(d).
	\label{eq:charge}
	\end{aligned}
\end{equation}
All parameters and details of their physical meanings should be referred to the original literature. Since all the parameters of the above equation are constants, one can simplify them as
\begin{equation}
	\begin{aligned}
		&C_{pp}^{s=0}=C_{np}^{s=0}=k_1\frac{Z^2}A,\\
		&C_{pn}^{s=1}=k_2A,
		\label{eq:charge2}
	\end{aligned}
\end{equation}
where $k_1\simeq0.023$ and $k_2\simeq0.085$ are constants come from the parameters in Eq. (\ref{eq:charge}). If we consider the symmetric nuclei, $Z=A/2$, and $C_{pp}/C_{np}=c_{pp}/c_{np}$ with $c_{ij}=C_{ij}/\frac{A}{2}$, one should conclude
\begin{equation}
	\frac{c_{pp}}{c_{np}}=\frac{\frac{1}{4}k_1}{\frac{1}{4}k_1+k_2}\simeq0.06.
	\label{eq:charge3}
\end{equation}
This leads to a constant value for the ratio $R$ in Eq. (\ref{eq:ratio}) if the reduced nuclear contacts do not depend on the nucleon number $A$. This is not consistent with the results in Ref. \cite{Weiss:2016obx}. Although there is not enough evidence to say whether the reduced contacts are $A$-dependent, for the time being we identify the extraction results in Ref. \cite{Weiss:2016obx}. As an exercise, we just keep the proportionality of Eq. (\ref{eq:charge2}) but do not force the numerator and denominator to be constants.

\begin{figure}[htbp]
	\centering
	\includegraphics[width=0.46\textwidth]{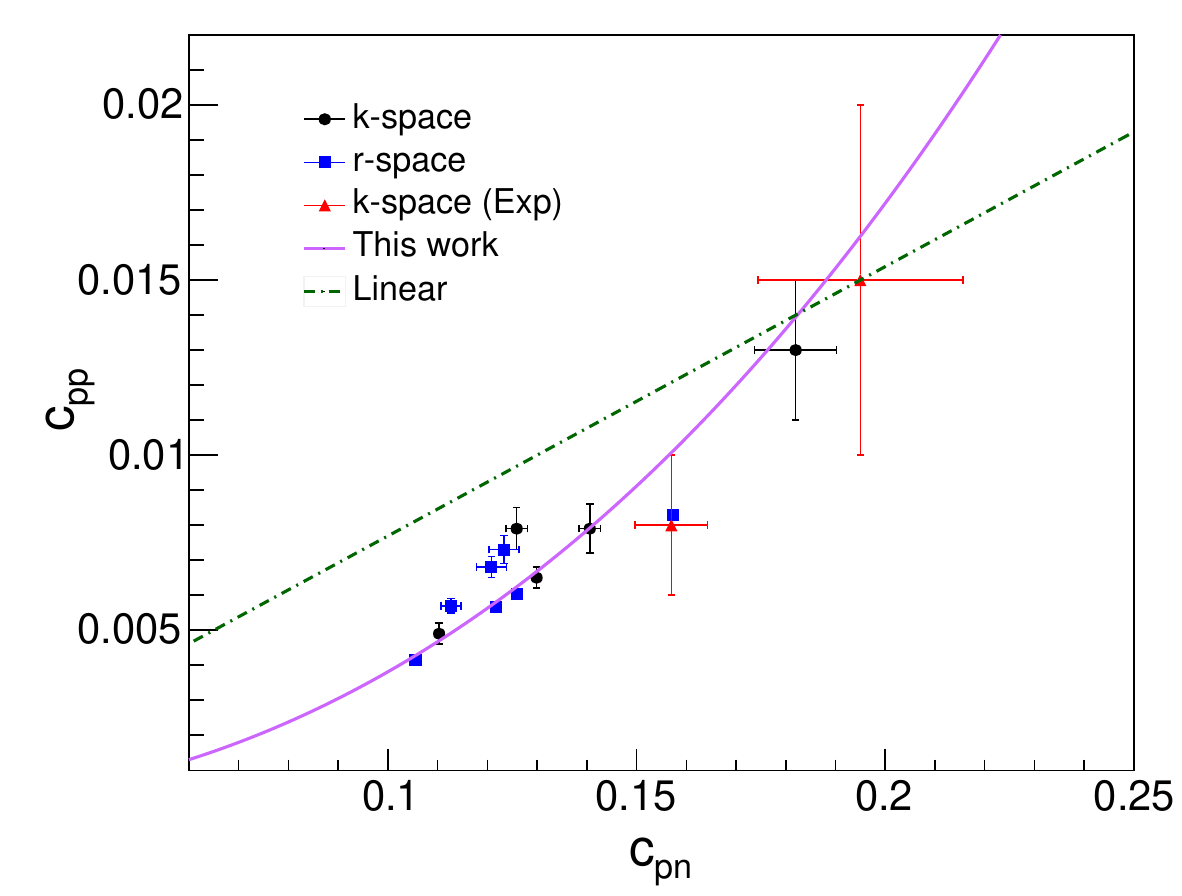}
	\caption{The reduced nuclear contacts with two channels $pp$ and $pp$. All data points were extracted in Ref. \cite{Weiss:2016obx}. The green dashed-dotted line shows the result in Eq. (\ref{eq:charge3}) and we do not force the numerator and denominator to be constants. The violet line represents $R=2$ in Eq. (\ref{eq:ratio}), we do not have the formula which gives us the exact relation of $c_{pp}$ and $c_{np}$.}
	\label{fig:compare}
\end{figure}

As another possibility, although we have not mathematically found the hidden relation that makes $R$ a constant, the results (see below) show that they can reasonably well correspond to the currently extracted nuclear contacts. Figure. \ref{fig:compare} presents a comparison of the results of the above two possibilities. The black and blue data points come from Ref. \cite{Weiss:2016obx} in $k$-space and $r$-space distribution fitting, while the red data come from Refs. \cite{Subedi:2008zz,LabHallA:2014wqo,Hen:2014nza,JeffersonLabHallA:2007lly,Weiss:2016obx}. Note that the only symmetric nuclei case is considered. The green dashed line comes from Eq. (\ref{eq:charge2}). From Eq. (\ref{eq:charge3}) it can be seen that when we consider reduced nuclear contacts $c_{pp}$ and $c_{np}$ are determined by the parameters $k_1$ and $k_2$ respectively, i.e. there is only one point in the $c_{pp}-c_{np}$ plane. Based on the current extracted nuclear contacts results we have no reason to think that this is reasonable. Thus we only consider proportionality conclusions but do not insist on whether nuclear contacts are constant or not.

Let's try to discuss the main points of this section. First, based on the nuclear contacts shown in Ref. \cite{Weiss:2016obx} we compute the SRC orbital entanglement entropy for different channels and obtain a ratio relation as mentioned in the text, which is denoted by Eq. (\ref{eq:ratio}) The main results are shown in Table. \ref{tab:contacts} and Figures. \ref{fig: ratio-B}-\ref{fig: ratio-A}. Second, we attempt to understand what causes this fixed ratio. In principle, making $R\simeq2.0$ in the joint action of Eqs. (\ref{eq:charge}-\ref{eq:charge3}) one also needs to consider that the reduced nuclear contacts do not depend on the number of nucleons $A$. Furthermore, although the exact reason why $R$ converges to 2 in any current nuclei is not yet known, the introduction of entanglement entropy gives an additional constraint on the value of nuclear contacts (see the violet line in Figure. \ref{fig:compare}).

\section{Conclusion and outlook}
\label{sec:summary}

In this work, we use the nuclear contacts to characterize the effects of nuclear SRC from GCF. We also compute implicit constraints on the values of nuclear contracts using the single-orbital SRC entanglement entropy method. On this basis, we currently consider that the ratio of SRC entanglement entropy to the corresponding nuclear contacts for different channels does not depend on the nucleon number of the atomic nuclei (Figure. \ref{fig: ratio-A}), and the ratio is close to the constant 2. We mainly discuss the symmetric nuclei case in the paper. For asymmetric nuclei case, it has little impact on the main results. We suppose therefore the findings are universal for any symmetric nuclei.

It should be noted that we only consider the single-orbital SRC entanglement entropy computational model, and for the two-orbital case, one can refer to Ref. \cite{Robin:2020aeh} for some details. Our result is an extension of Ref. \cite{Pazy:2022mmg} and suggests that the introduction of entanglement entropy can constrain the value of nuclear contacts taken in the GCF approach. Of course, our results are based on GCF and are model-dependent. The high-precision testing of the model relies on future high-precision SRC measurements. In fact, the contribution of the wave function of high-excited single-nucleon or two-nucleon systems to SRC is also an interesting issue. On this basis, the study of entanglement entropy requires us to deepen our understanding of nuclear contacts.

Our results give a continuum of nuclear contacts assignments for different channels and are likely to apply to all symmetric nuclei. Since nuclear contacts in GCF can characterize the ratio of SRC nucleon pairs to the nuclear mass number, it becomes a possibility that our results provide a prediction for future work on measuring the ratio of SRC. In principle, the conclusions we give can be used when determining the specific information of an SRC channel ($np$) to get the SRC information of another channel ($pp$).

\begin{acknowledgments}
This work is supported by the Strategic Priority Research Program of Chinese Academy of Sciences (Grant NO. XDB34030301).
\end{acknowledgments}

\bibliographystyle{apsrev4-1}
\bibliography{refs}
\newpage

\end{document}